\documentclass[journal,onecolumn,12pt]{IEEEtran}
\usepackage{amssymb}
\usepackage{amsmath}
\usepackage{graphicx}
\usepackage{tcolorbox}

\usepackage[ruled]{algorithm2e}
\usepackage{bm}
\usepackage{romannum}
\usepackage{xcolor}
\usepackage{enumitem}
\usepackage{hyperref}
\hypersetup{
    colorlinks=true,
    linkcolor=black,
    filecolor=black,      
    urlcolor=black,
}

\makeatletter
\newcommand{\subalign}[1]{%
  \vcenter{%
    \Let@ \restore@math@cr \default@tag
    \baselineskip\fontdimen10 \scriptfont\tw@
    \advance\baselineskip\fontdimen12 \scriptfont\tw@
    \lineskip\thr@@\fontdimen8 \scriptfont\thr@@
    \lineskiplimit\lineskip
    \ialign{\hfil$\m@th\scriptstyle##$&$\m@th\scriptstyle{}##$\hfil\crcr
      #1\crcr
    }%
  }%
}
\makeatother

\newcommand{\norm}[1]{\left\lVert#1\right\rVert}

\DeclareMathOperator*{\argmax}{arg\,max}

\usepackage{siunitx}
\begin{document}
\title{Meta-Learning to Communicate:\\ Fast End-to-End Training for Fading Channels}
\author{Sangwoo Park,~\IEEEmembership{Student Member,~IEEE,}
        Osvaldo Simeone,~\IEEEmembership{Fellow,~IEEE,}
        and\\~Joonhyuk Kang,~\IEEEmembership{Member,~IEEE}%

\thanks{Code for regenerating the results of this paper can be found at https://github.com/kclip/meta-autoencoder.}
\thanks{The work of S. Park and J. Kang was supported by the National Research Foundation of Korea (NRF) grant funded by the Korea government (MSIT) (No. 2017R1A2B2012698). The work of O. Simeone was supported by the European Research Council (ERC) under the European Union's Horizon 2020 research and innovation programme (grant agreement No. 725731).}}


\pagenumbering{arabic}

\maketitle
\thispagestyle{plain}
\pagestyle{plain}

\begin{abstract}
When a channel model is available, learning how to communicate on fading noisy channels can be formulated as the (unsupervised) training of an autoencoder consisting of the cascade of encoder, channel, and decoder. An important limitation of the approach is that training should be generally carried out from scratch for each new channel. To cope with this problem, prior works considered joint training over multiple channels with the aim of finding a single pair of encoder and decoder that works well on a class of channels. As a result, joint training ideally mimics the operation of non-coherent transmission schemes. In this paper, we propose to obviate the limitations of joint training via meta-learning: Rather than training a common model for all channels, meta-learning finds a common initialization vector that enables fast training on any channel. The approach is validated via numerical results, demonstrating significant training speed-ups, with effective encoders and decoders obtained with as little as one iteration of Stochastic Gradient Descent.
\end{abstract}



%
\IEEEpeerreviewmaketitle

\vspace{0.2cm}
\section{Introduction}
\label{sec:intro}

\begin{figure}[t!]
    \centering
    \includegraphics[width=0.8\columnwidth]{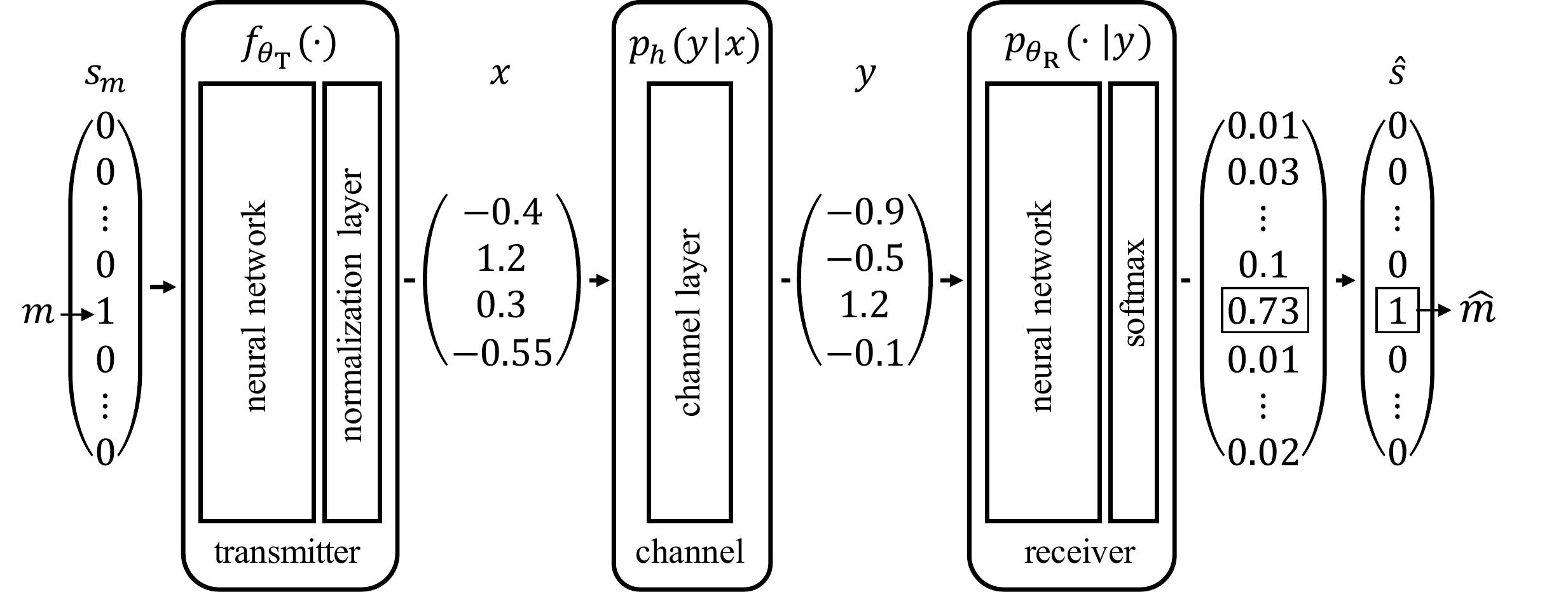}
    \caption{Single link modeled as an autoencoder: A message $m$ is mapped into a codeword $x$ via a trainable encoder $f_{\theta_\text{T}}(\cdot)$, while the received signal $y$, determined by the channel $p_{h}(y|x)$, is mapped into estimated message $\hat{m}$ through a trainable decoder $p_{\theta_\text{R}}(\cdot|y)$.}
    \label{fig:model}
\end{figure}

There is current renewed interest in the idea of substituting the conventional model-based design flow of communication systems with data-driven approaches. Among the key advantages of data-driven designs are the facts that they can be automated, yielding solutions of potentially lower design and implementation complexity; and that they do not require the availability of analytically tractable channel models \cite{o2017introduction, simeone2018very}. A fundamental model for the end-to-end design of a communication link views a communication link as an autoencoder consisting of the cascade of encoder, channel, and decoder (see Fig.~\ref{fig:model}) \cite{o2017introduction}. Autoencoders can be trained in an unsupervised way with the aim of reproducing the input, here the communication message, to the output, here the decoded message \cite{simeone2018brief}. If a channel model is available to draw samples of channel outputs given the inputs, Stochastic Gradient Descent (SGD) can be used to train simultaneously both encoder and decoder \cite{o2017introduction}.

The original work \cite{o2017introduction} has been extended in a number of directions. Notably, in \cite{dorner2017deep} a practical implementation is showcased that accounts for imperfect synchronization and other hardware impairments. In \cite{bourtsoulatze2019deep}, the approach is used to train a system that carries out joint source-channel coding for video transmission.  In \cite{aoudia2019model}, the authors propose a solution that overcomes the need for channel models by training the encoder via reinforcement learning on the basis of feedback from the receiver. More discussion on the state of the art can be found in \cite{simeone2018very}. 

An important limitation of the autoencoder-based approach to end-to-end training is that training should be generally carried out from scratch for each new channel. To cope with this problem, prior works considered joint training over multiple channels with the aim of finding a single pair of encoder and decoder that works well on a class of channels \cite{o2017introduction, simeone2018very}. As a result, joint training ideally mimics the operation of non-coherent transmission schemes. In this paper, we propose to obviate the limitations of joint training via meta-learning: Rather than training a common model for all channels, meta-learning finds a common initialization vector that enables fast training on any channel. 

Meta-learning is currently under intensive investigation in the field of machine learning, with one of the most notable new contributions given by the introduction of the gradient-based scheme Model-Agnostic Meta-Learning (MAML) \cite{finn2017model}. Applications of meta-learning to communication systems are currently sparse: In our prior work \cite{park2019learning}, we have applied meta-learning to the problem of detection; paper \cite{jiang2019mind} considered decoding; and \cite{mao2019roemnet} studied channel estimation.  
\vspace{0.2cm}
\section{System Model and Background}
\label{sec:model}

We consider the single-link set-up of \cite{o2017introduction}, in which the goal is to train in an end-to-end fashion encoder and decoder of a communication system based on the availability of a channel model. For this purpose, as illustrated in Fig.~\ref{fig:model}, the cascade of coder, channel, and decoder is modeled as an autoencoder with trainable weights $\theta=(\theta_\text{T},\theta_\text{R})$ describing the operation of encoder ($\theta_\text{T}$) and decoder ($\theta_\text{R}$).

The encoder takes as input a one-hot vector $s_m$ of dimension $2^k$, which represents a message $m\in\{1,\ldots,2^k\}$ of $k$ bits. Vector $s_m$ has a single entry equal to ``1'' in position $m$, with all other entries equal to zero. The encoder maps each input $s_m$ into a transmitted vector $x\in\mathbb{C}^n$ of $n$ complex symbols or equivalently, $2n$ real symbols. As seen in Fig.~\ref{fig:model}, the encoding from $s_m$ to $x$ is done through a trainable mapping $x=f_{\theta_\text{T}}(s_m)$, which is defined by a neural network with weight vector $\theta_\text{T}$ and by a normalization layer that ensures the total power constraint $\norm{x}^2/n=E_s$.

The codeword $x$ is transmitted through a channel $p_h(y|x)$, which is not necessarily memoryless, to produce the received signal $y$. We will specifically assume the general form 
\begin{align} \label{eq:channel_model}
y = h*x + w,
\end{align}
where $w\sim\mathcal{CN}(0,N_0)$ represents complex Gaussian i.i.d. noise and ``$*$'' indicates a linear operation on input $x$ parameterized by a channel vector $h$, e.g., a convolution. 

The receiver passes the received signal through a neural network parameterized by a weight vector $\theta_\text{R}$ that outputs a $2^k\times 1$ vector of probabilities $p_{\theta_\text{R}}(m|y)$ for $m\in\{ 1,\ldots,2^k \}$. Each output $p_{\theta_\text{R}}(m|y)$ provides an estimate of the corresponding posterior probability that the transmitted message is $m$. A final hard estimate can be obtained via the approximate maximum a posteriori (MAP) rule
$\hat{m} =\argmax_{m=1,\ldots,2^k} p_{\theta_\text{R}}(m|y).$

As a surrogate of the probability of error $\textrm{P}_e = \Pr[\hat{m}\neq m]$, for a given channel state $h$, a typical choice is the cross-entropy (or log-) loss (see, e.g., \cite{o2017introduction, simeone2018very}) 
\begin{align} \label{eq:ce}
L_h(\theta) = -\mathbb{E}_{\subalign{m&\sim p(m),\\y&\sim p_h(y|f_{\theta_\text{T}}(s_m))}}[\log p_{\theta_\text{R}}(m|y)],
\end{align}
where the message probability distribution is typically taken to be uniform as $p(m)=2^{-k}$. The parameter vector $\theta$ can be trained by maximizing \eqref{eq:ce} via SGD \cite{o2017introduction}. To this end, at each iteration $i$, one draws $P$ independent samples $\mathcal{D}^{(i)}=\{ m_j\sim p(m), w_j \sim \mathcal{CN}(0,N_0) \}_{j=1}^P$ and then uses gradient descent on the empirical cross-entropy loss
\begin{align} \label{eq:empirical_loss}
\hat{L}_h(\theta) = -\frac{1}{P} \sum_{j=1}^P \log p_{\theta_\text{R}}(m_j|h*f_{\theta_\text{T}}(s_{m_j})+w_j).
\end{align}
This yields the gradient update rule \eqref{eq:sgd} for some learning rate $\eta > 0$. Note that the gradient in \eqref{eq:sgd} with respect to $\theta_\text{T}$ can be interpreted as an example of pathwise gradient estimators (or of the reparameterisation trick) \cite{mohamed2019monte}. 
This training procedure is summarized for reference in Algorithm~\ref{alg:simple_ae}.

\begin{algorithm}[h] 
\DontPrintSemicolon
\smallskip
\KwIn{channel $h$; number of samples per iteration $P$; step size hyperparameter $\eta$}
\KwOut{learned parameter vector $\theta$}
\vspace{0.15cm}
\hrule
\vspace{0.15cm}
{\bf initialize} $\theta^{(0)}$ \\
$i = 0$ \\
\While{{\em not converged}}{

\vspace{0.2cm}
draw $P$ i.i.d. samples 

\vspace{0.1cm}\hspace{0.05cm}
$\mathcal{D}^{(i)}=\{ m_j\sim p(m), w_j \sim \mathcal{CN}(0,N_0) \}_{j=1}^P$\\
\vspace{0.1cm}
update autoencoder 

\vspace{-0.65cm} 
\begin{align} \label{eq:sgd}
\theta^{(i+1)} \leftarrow \theta^{(i)} - \eta \nabla \hat{L}_h(\theta^{(i)})
\end{align}\\
\vspace{-0.35cm}
$i\leftarrow i+1$
}
$\theta \leftarrow \theta^{(i)}$ 
\vspace{0.15cm}
\caption{Autoencoder-Based Learning}
\label{alg:simple_ae}
\end{algorithm} 
\vspace{0.2cm}
\vspace{-0.52cm}

\section{Fast Training via Meta-Learning} 
\label{sec:meta}

In practice, one is interested in designing encoder parameter $\theta_\text{T}$ and decoder parameter $\theta_\text{R}$ for any new channel $h$ by using a small number of iterations. In this section, we tackle this problem via meta-learning. The key idea is to choose the initialization $\theta^{(0)}$ in Algorithm~\ref{alg:simple_ae} in such a way that a few steps of SGD \eqref{eq:sgd} can yield effective encoder and decoder for \emph{any} channel $h$.

\subsection{Joint Learning}
\label{subsec:meta_setup}
In order to introduce meta-learning, let us first consider the simpler approach of \emph{joint training}. Under joint training, we wish to find a unique solution parameter $\theta$ that works well on all channels in a set $\mathcal{H}=\{ h_k \}_{k=1}^K$ sampled from a given fading distribution. The common model should hence ideally approximate some form of non-coherent transmission in order to work well on all channels in $\mathcal{H}$ (see, e.g., \cite{o2017introduction}). 

Mathematically, we can formulate the problem as the optimization of the sum log-loss \eqref{eq:ce} over the channels in $\mathcal{H}$
\begin{align} \label{eq:joint}
\min_\theta \sum_{h\in\mathcal{H}} L_{h}(\theta).
\end{align}
In a manner similar to \eqref{eq:sgd}, problem \eqref{eq:joint} can be tackled via SGD as
\begin{align} \label{eq:joint_sgd}
\theta^{(i+1)} \leftarrow \theta^{(i)} - \kappa \nabla \sum_{h\in\mathcal{H}} \hat{L}_{h}(\theta^{(i)}),
\end{align}
for iterations $i=1,2,\ldots$, where $\kappa>0$ is a learning rate and the function $\hat{L}_{h}(\theta^{(i)})$ is computed as in \eqref{eq:empirical_loss}.

\subsection{Meta-Learning}
\label{subsec:maml}

\begin{figure}[t!]
    \centering
    \includegraphics[width=0.7\columnwidth]{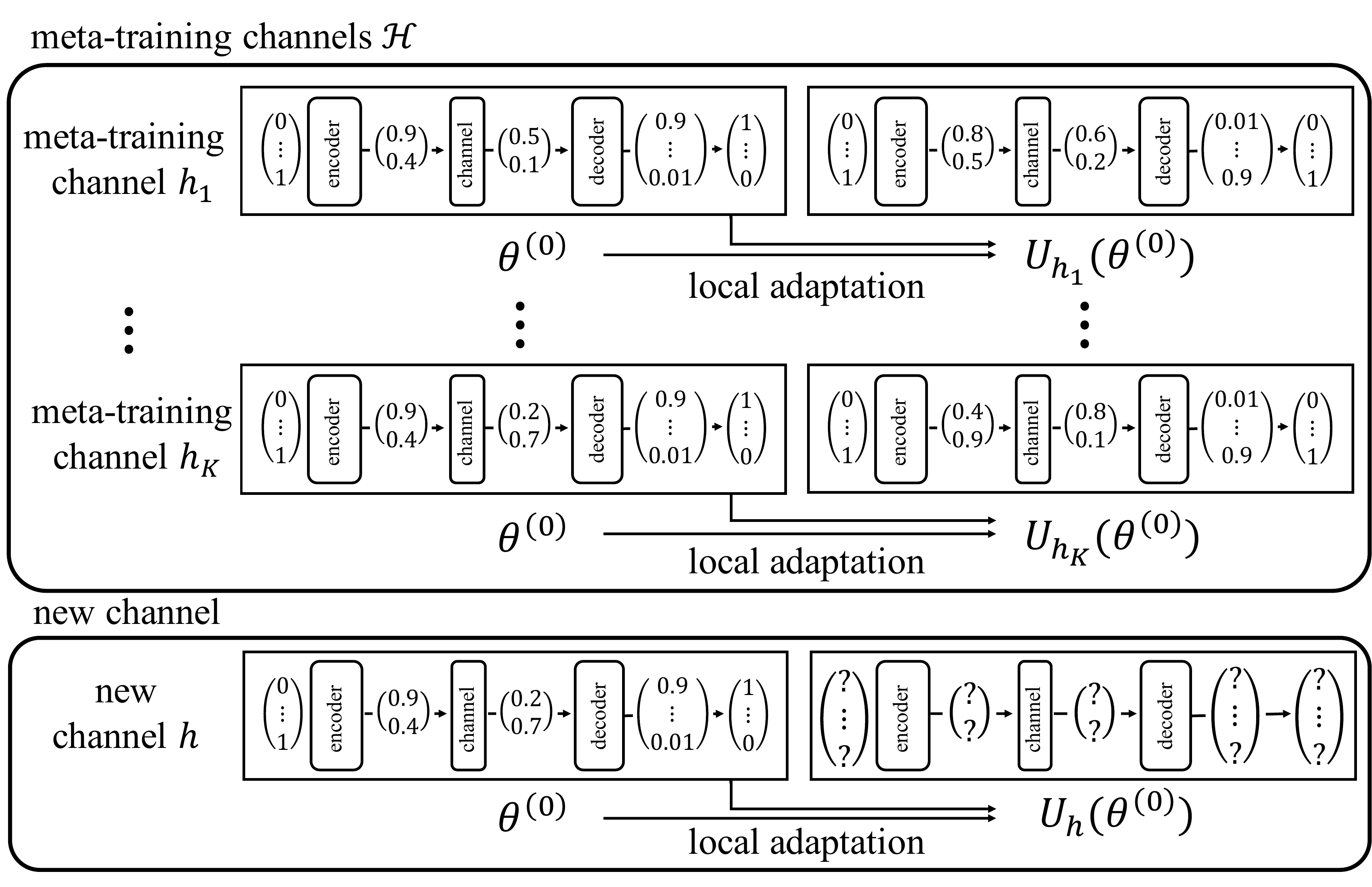}
    \caption{Meta-learning set-up via MAML: The goal is to find an initialization $\theta^{(0)}$ such that encoder and decoder parameters $\theta$ can be adapted to any channel with few SGD steps.}
    \label{fig:meta}
\end{figure}

While joint training obtains a common model $\theta$ for all channels in $\mathcal{H}$, meta-learning obtains a common initialization $\theta^{(0)}$, to be used for \emph{local training} as in Algorithm~\ref{alg:simple_ae}, for all channels in $\mathcal{H}$. As illustrated in Fig.~\ref{fig:meta}, the goal is to enable faster convergence, based on fewer iterations in Algorithm~\ref{alg:simple_ae}, to an effective solution for any channel related to those in $\mathcal{H}$. Therefore, ideally, meta-learning produces coherent encoding and decoding schemes more quickly, rather than providing optimized non-coherent solutions. 

As a notable example of meta-learning, MAML \cite{finn2017model} seeks for an initialization value $\theta^{(0)}$ that optimizes the problem
\begin{align} \label{eq:maml}
\min_{\theta^{(0)}}{L^\text{ML}} (\theta^{(0)}) = \sum_{h\in\mathcal{H}} L_{h}(\theta^{(0)}-\eta\nabla\hat{L}_{h}(\theta^{(0)})),
\end{align}
where the log-loss function $L_{h}(\cdot)$ and $\hat{L}_{h}(\cdot)$ are defined as in \eqref{eq:ce} and \eqref{eq:empirical_loss}, respectively. According to \eqref{eq:maml}, MAML aims to find an initialization value $\theta^{(0)}$ from which a single SGD update \eqref{eq:sgd} on each channel $h$ yields minimal log-loss. The key idea is that, if we succeed in finding an initialization point $\theta^{(0)}$ that yields a good solution for the channels in $\mathcal{H}$ after one SGD iteration, we can expect the initialization point $\theta^{(0)}$ to produce an effective autoencoder also for a new channel with only a small number of SGD iterations. 

Denoting as $U_h(\theta^{(0)}) = \theta^{(0)}-\eta\nabla\hat{L}_{h}(\theta^{(0)})$ the local updated vector for channel $h$, MAML tackles \eqref{eq:maml} via SGD through the update \eqref{eq:gd_ML}, which can be computed as
\begin{align} \label{eq:meta_update}
\nonumber
\theta^{(0, i+1)} &\leftarrow \theta^{(0,i)} - \kappa\sum_{h\in\mathcal{H}}(\mathbf{J}_{\theta^{(0)}}U_h(\theta^{(0,i)}))\nabla L_{h}(U_h(\theta^{(0,i)}))
\\
& \approx  \theta^{(0,i)} - \kappa \sum_{h\in\mathcal{H}} (I-\eta\nabla^2 \hat{L}_{h}(\theta^{(0,i)})) \nabla\hat{L}'_{h}(U_h(\theta^{(0,i)})),
\end{align}
for iterations $i=1,2,\ldots$, where $\mathbf{J}_{\theta^{(0)}}$ represents the Jacobian operation. The approximation in the last line is due to the estimation of the cross-entropy loss $L_{h}(\cdot)$ with the empirical loss $\hat{L}'_{h}(\cdot)$ obtained using $P$ i.i.d. samples $\mathcal{D}^{(i)'}=\{ m_j\sim p(m), w_j \sim \mathcal{CN}(0,N_0) \}_{j=1}^P$. The full algorithm is summarized in Algorithm~\ref{alg:ae_training}.

\begin{algorithm}[h] 
\DontPrintSemicolon
\smallskip
\KwIn{meta-training channels $\mathcal{H} = \{ h_k \}_{k=1}^K$; number of samples per iteration $P$; step size hyperparameters $\eta$ and $\kappa$}
\KwOut{learned initial parameter vector $\theta^{(0)}$}
\vspace{0.15cm}
\hrule
\vspace{0.15cm}
{\bf initialize} parameter vector $\theta^{(0,0)}$ \\
$i=0$ \\
\While{{\em not converged}}{
\For{{\em each meta-training channel $h\in\mathcal{H}$}}{
\vspace{0.05cm} compute local adaptation $U_h(\theta^{(0,i)}) = \theta^{(0,i)}-\eta\nabla\hat{L}_{h}(\theta^{(0,i)})$\\
}
update initialization parameter vector 
\vspace{-0.25cm}
\begin{align} \label{eq:gd_ML}
\theta^{(0,i+1)} \leftarrow \theta^{(0,i)} - \kappa \nabla L^\text{ML}(\theta^{(0,i)})
\end{align}
\\
\vspace{-0.4cm}
$i\leftarrow i+1$
}
\caption{Autoencoder-Based Meta-Learning}
\label{alg:ae_training}
\end{algorithm}  
\vspace{0.2cm}
\section{Experiments and Final Remarks}
\label{sec:exp}

\subsection{Toy Example}
\label{subsec:exp-toy}

Before considering realistic fading channels, we start with a simple example in which we transmit $k=2$ bits over $n=1$ complex symbol. We consider a simple channel model $p_{h}(\cdot|x) = h x + w$, where channel state $h$ has fixed amplitude $|h| = 1$ with two possible phases $\measuredangle{h} \in \{ \frac{\pi}{4}, \frac{3\pi}{4} \} $. The signal-to-noise ratio (SNR) is given as $E_s/N_0= 2E_b/N_0$ with $E_b/N_0=15\text{ dB}$. The phase of the new channel state is selected randomly between $\pi/4$ and $3\pi/4$ with equal probability, while the meta-learning set is given as $\mathcal{H} = \{ e^{\pi/4}, e^{3\pi/4} \}$. We set $P=4$ samples per iteration in \eqref{eq:empirical_loss}. We compare the performance of the proposed meta-learning approach with: (\emph{i}) \emph{fixed initialization}, in which the initialization vector $\theta^{(0)}$ is randomly chosen; and (\emph{ii}) \emph{joint training}. For (\emph{ii}), we use learning rate $\kappa=0.01$ in \eqref{eq:joint_sgd} with Adam \cite{kingma2014adam}, and then, for the new channel, a learning rate $\eta=0.001$ with Adam in Algorithm~\ref{alg:simple_ae}. For meta-training, we used learning rates $\eta=0.1$ and $\kappa=0.01$ and Adam in Algorithm~\ref{alg:ae_training}. For the new channel, we used learning rate $\eta = 0.1$ for one iteration, and then Adam with $\eta=0.001$ in Algorithm~\ref{alg:simple_ae}.

The encoder $f_{\theta_\text{T}}(\cdot)$ is a neural network with three layers, i.e., an input layer with $2^k=4$ neurons, one hidden layer with four neurons with rectified linear unit (ReLU) activation, and one output layer with $2n=2$ linear neurons, each encoding a real or imaginary component of $x$, followed by the normalization layer. The decoder $p_{\theta_\text{R}}(\cdot|y)$ is also a neural network with three layers, i.e., an input layer with $2n=2$ neurons, one hidden layer composed of four neurons with ReLU activation, and one softmax output layer with $2^k=4$ neurons. In addition to this \emph{vanilla} autoencoder, illustrated in Fig.~\ref{fig:model}, we consider the more sophisticated receiver design proposed in \cite{o2017introduction} which adds a Radio Transformer Networks (RTN) to the receiver. The RTN applies a filter $w$ to the received signal $y$ to obtain the input $\bar{y} = y * w$ to the decoder as $p_{\theta_\text{R}}(\cdot|\bar{y})$. The filter $w$ is obtained at the output of an additional neural network, which has four layers, i.e., an input layer with $2n=2$ neurons, two hidden layers each with two neurons with hyperbolic tangent function as activation function, and one output layer with two linear neurons.

\begin{figure}[t]
    \flushleft
    \hspace*{0.02in}
    \centering
    \includegraphics[width=0.6
    \columnwidth]{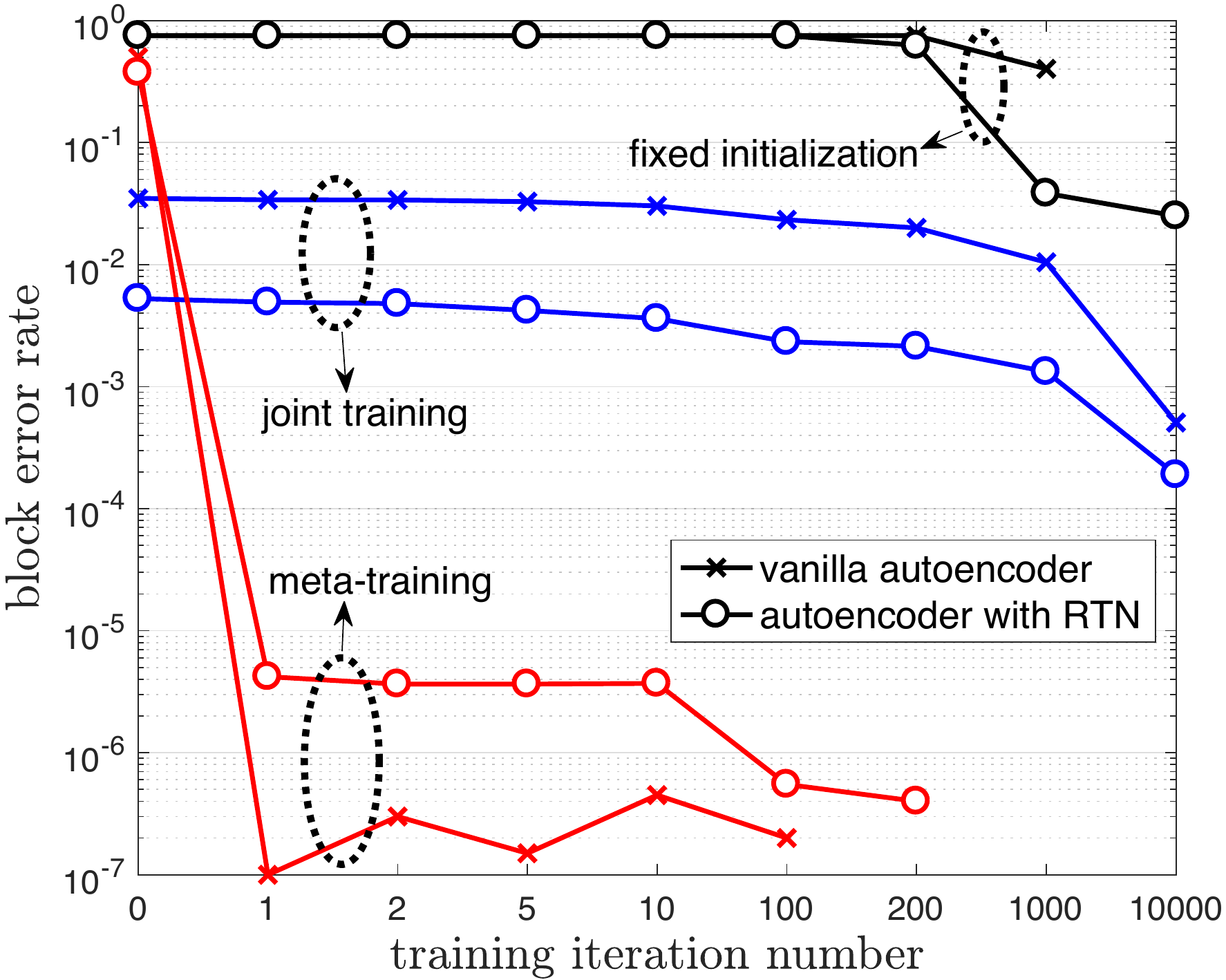}
    \caption{Block error rate over iteration number for training on the new channel ($k=2$ bits, $n=1$ complex channel uses, two possible channels with phase $\pi/4, 3\pi/4$, $P=4$ messages per iteration).}
    \label{fig:toy_1}
\end{figure}

\begin{figure}[t]
    \flushleft
    \hspace*{0.02in}
    \centering
    \includegraphics[width=0.7
    \columnwidth]{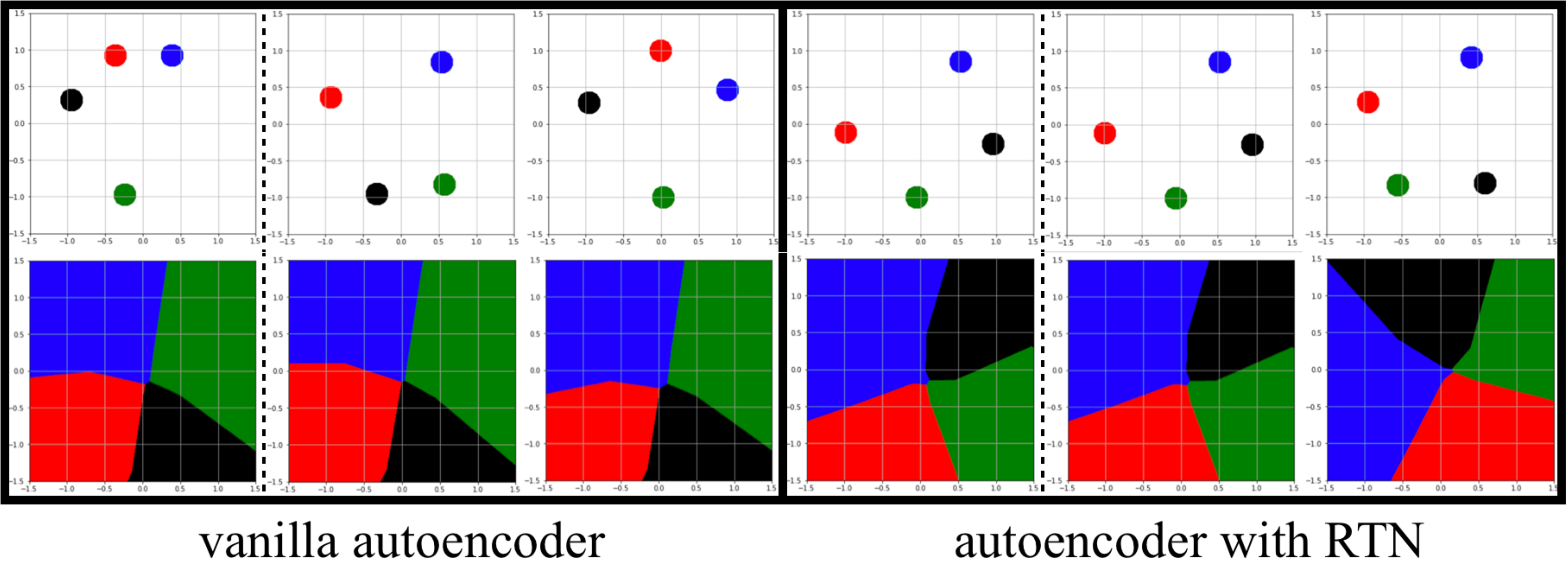}
    \caption{Ilustration of (top) constellation points, (bottom) decision region; (left) before adaptation, (center) after one SGD step for channel with phase $\pi/4$, (right) after one SGD step adaptation for channel with phase $3\pi/4$. Best viewed in color printed version.}
    \label{fig:toy_vis}
\end{figure}

In Fig.~\ref{fig:toy_1}, we plot the average block error rate with respect to number of iterations for training on the new channel. We averaged over $20$ new channels and $10^6$ messages per each channel. Meta-learning is seen to offer much faster convergence than the mentioned baseline approaches, effectively adapting to the new channel even with only one SGD iteration. In contrast, joint training shows small improvements unless the number of iterations on the new channel is large enough. It is also worth noting that except for joint training, the vanilla autoencoder architecture outperforms the autoencoder architecture with RTN. This is due to the larger number of parameters in the RTN model, which are generally difficult to train with few iterations. 

In order to gain intuition on how meta-learning enables faster learning, in Fig.~\ref{fig:toy_vis}, we plot the constellation points produced by encoder $f_{\theta_\text{T}}(\cdot)$ (top) and decision regions through $p_{\theta_\text{R}}(\cdot|y)$ (bottom). For both autoencoder architectures, the leftmost column refers to the situation before adaptation to a new channel; the center column refers to the situation after one iteration on the new channel with phase $\pi/4$; and the rightmost column corresponds to the situation after one iteration for the new channel with phase $3\pi/4$. Meta-learning is seen to provide an initialization that works well on one channel while adapting to the other channel after a single iteration. Furthermore, the vanilla autoencoder is seen to adapt only the encoder, while the autoencoder architecture with RTN only adapts the decoder. This suggests that, for the autoencoder with RTN, it is easier to adapt the decoder via changes to the RTN module.

\subsection{A More Realistic Scenario}
\label{sec:exp-real}

We now consider a more realistic scenario including Rayleigh fading channels. To this end, we transmit $k=4$ bits over $n=4$ complex symbols of a frequency-selective Rayleigh block fading channels with $L=3$ taps. The taps are independently generated as $\mathcal{CN}(0, L^{-1})$ variables, while the SNR is $E_s/N_0= 15\text{ dB}$. The number of meta-training channels is $K=100$ and the number of samples per iteration is $P=16$, including all $2^k=16$ messages per iteration while the RTN filter has $2L=6$ taps.

\begin{figure}[t]
    \flushleft
    \hspace*{0.02in}
    \centering
    \includegraphics[width=0.6
    \columnwidth]{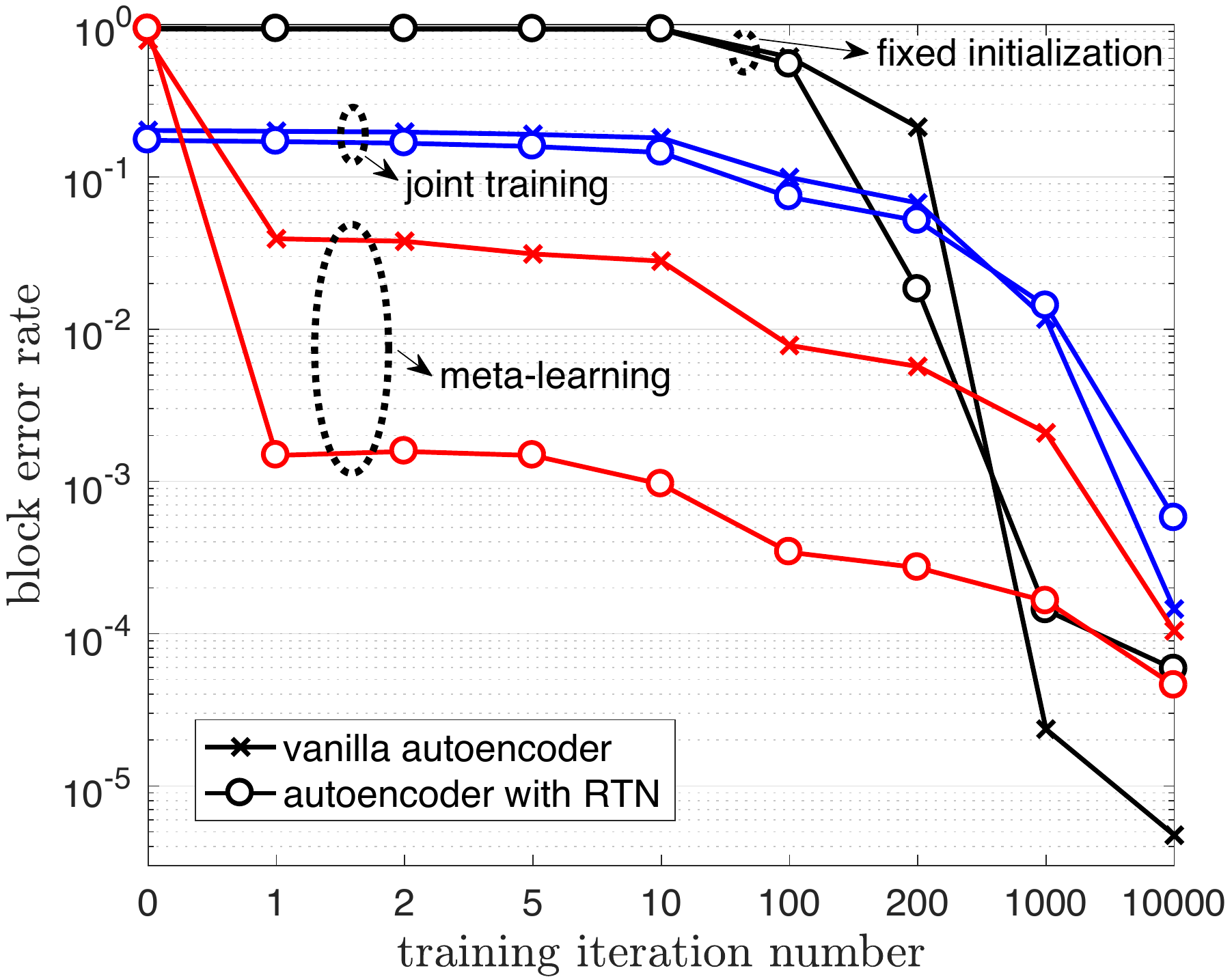}
    \caption{Block error rate over iteration number for training on the new channel ($k=4$ bits, $n=4$ complex channel uses, Rayleigh block fading channel model with $L=3$ taps, $P=16$ messages per iteration).}
    \label{fig:real}
\end{figure}

Confirming the results for the toy example, Fig.~\ref{fig:real} shows that meta-learning enables faster convergence than joint training. Unlike for the toy example, for both joint training and meta-learning, RTN provides performance benefits due to the more challenging propagation environment. Furthermore, with extremely long training, e.g., $10^3-10^4$ iterations, fixed initialization outperforms all other schemes suggesting that the inductive bias adopted by joint and meta-learning may cause performance degradation when sufficient data is available for a channel.
  
\vspace{0.2cm}

\bibliographystyle{IEEEtran}
\bibliography{ref}

\end{document}